# Improving Effectiveness Of E-Learning In Maintenance Using Interactive-3D

Lt.Dr.S Santhosh Baboo

Reader

P.G. & Research Dept of Computer Science

D.G.Vaishnav College, Chennai 106

Nikhil Lobo

Research Scholar

Bharathiar University

nikhillobo@baehal.com

*Abstract*—In aerospace and defense, training is being carried out on the web by viewing PowerPoint presentations, manuals and videos that are limited in their ability to convey information to the technician. Interactive training in the form of 3D is a more cost effective approach compared to creation of physical simulations and mockups. This paper demonstrates how training using interactive 3D simulations in e-learning achieves a reduction in the time spent in training and improves the efficiency of a trainee performing the installation or removal.

## I. INTRODUCTION

Some procedures are found to be hazardous and need to be demonstrated to maintenance personnel without damaging equipment or injuring personnel. These procedures require continuous practice and when necessary retraining.

The technician is also to be trained in problem solving and decision making skills. The training should consider technicians widely distributed with various skills and experience levels.

The aerospace and Defense industry in the past have imparted training using traditional blackboard outlines, physical demonstrations and video that is limited in their ability to convey information about tasks, procedures and internal components. Studies have demonstrated that 90% what they do is remembered by a student in contrast to 10% of what they read. Now a need has arisen for interactive three-dimensional content for visualization of objects, concepts and processes. Interactive training in the form of 3D is a more cost effective approach compared to creation of physical simulations and mockups. Online training generally takes only 60 percent of the time required for classroom training on the same material [1].

For any removal or installation procedure, it is important that the e-learning content should demonstrate the following aspects:

- What steps are to be carried out before starting a procedure?

- What special tools, consumables and spares are required for carrying out the procedure?

- What safety conditions are to be adhered to when carrying out the procedure?

- What steps a technician needs to perform as part of the procedure?

- What steps are to be carried out after completing a procedure?

## II. TRAINING EFFECTIVENESS

Effectiveness of training based on statistics convey that trainees generally remember more of what they see than of what they read or hear and more of what they hear, see and do than what they hear and see.

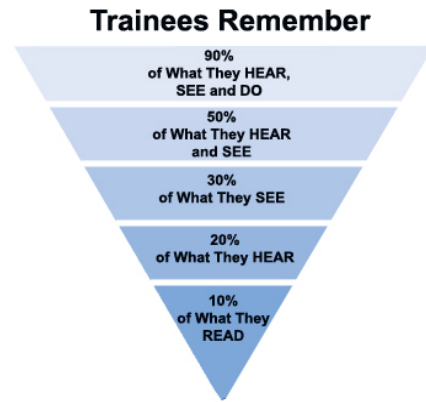

Figure 1. Statistics of training effectiveness

Three-dimensional models are widely used in the design and development of products because they efficiently represent complex shape information. These 3D models can be used in e learning to impart training. The training process can be greatly enhanced by allowing trainees to interact with these 3D models. Moreover by using WWW-based simulation, the company can make a single copy of the models available over the WWW instead of mailing demonstration software to potential customers. This reduces costs and avoids customer frustration with installation and potential hardware compatibility problems [2]. This simulation-based e-learning is designed to simplify and control reality by removing complex systems that exist in real life, so the learner can focus on the knowledge to be learnt effectively and efficiently [3].

The objectives of using interactive 3D for training is as follows:

- Reducing the time spent in training by 30% or more

- Reducing the time spent in performing the installation by a trainee by 25% or more

## III. COURSEWARE STRUCTURE

Since the web has been providing unprecedented flexibility and multimedia capability to deliver course materials, more and more courses are being delivered through the web. The existing course materials like PowerPoint presentations, manuals and videos, had a limited ability to convey information to the technician and most current internet-based educational applications do not present 3D objects even though 3D visualization is essential in teaching most engineering ideas. Interactive learning is essential for both acquiring knowledge and developing physical skills for carrying out maintenance related activities. Interaction and interactivity are fundamental to all dynamic systems, particularly those involving people [4]. Although there are images of three-dimensional graphics on the web, their two-dimensional format does not imitate the actual environment because objects are three-dimensional. Hence there is a need for integrating three-dimensional components and interactivity, creating three-dimensional visualization providing technicians an opportunity to learn through experimentation and research. The e-learning courseware is linearly structured with three modules for each removal or installation procedure. A technician is required to complete each module before proceeding to the next. These modules include Part Familiarization, Procedure and Practice. These modules provide a new and creative method for presenting removal or installation procedures effectively to technicians.

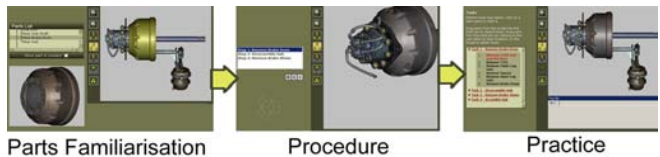

Figure 2.   Part Familiarization, Procedure and Practice modules

## IV. PART FAMILIARIZATION

This familiarizes the technician with the parts that constitute the assembly to be installed or removed. This module provides technicians with information as to what each part looks like and where they are located in the assembly. An assembly is displayed with a part list comprising of the parts that make up the assembly. Here the assembly is displayed as a 3D Model in the Main Window allowing the technician to rotate and move the assembly. Individual parts can be identified by selecting them from the parts list and by viewing the model in

"context view". Context View displays only the part selected in the part list with 100% opacity while reducing the opacity of other parts in the assembly. This enables easy identification of a selected part in the assembly. Individual parts that are selected are also displayed in a Secondary Window, allowing the technician to move or rotate the individual part for better understand of the particular part. Each part is labeled with a part number that is unique and a nomenclature.

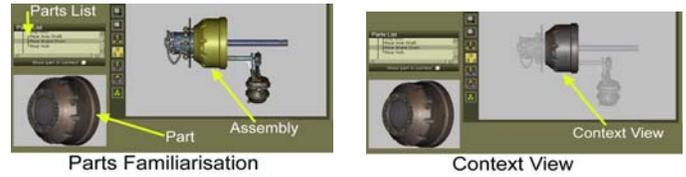

Figure 3.   Part Familiarisation module and Context View

## V. PROCEDURE

Once the technician has a clear understanding of the assembly and its parts, technicians advance to the module to learn how to accurately remove or install the assembly. In procedure, removal or installation is demonstrated in an animated form one step at a time. This allows the technician to study step-by-step removal or installation process using animation that technicians can replay anytime. The use of three-dimensional models in the animation imitates the real removal or installation process helping the technicians to understand concepts very clearly. Removal or installation of each part from the assembly is animated along with callouts indicating the part number, nomenclature, tool required and torque. The animations are presented one step at a time to ensure technicians are able to perform the removal or installation process in the right order. Safety conditions like warnings and cautions are also displayed along with the animation. A warning is used to alert the reader to possible hazards, which may cause loss of life, physical injury or ill health. A caution is used to denote a possibility of damage to material but not to personnel. A voice accompanies the animation to enable the technician to understand the procedure better.

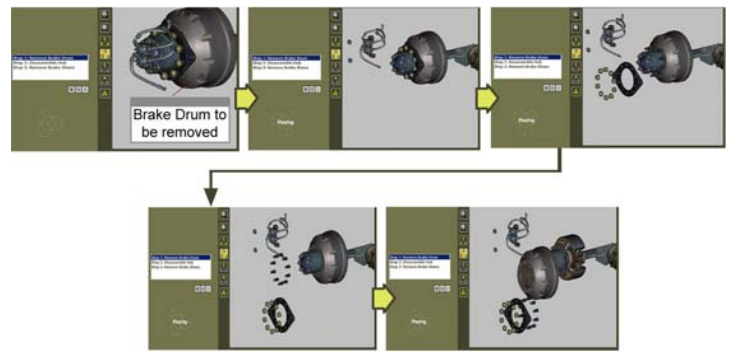

Figure 4.   Procedure  module

## VI. PRACTICE

Before performing any real procedure, technicians are first evaluated using a removal or installation procedural simulation. Practice allows a technician to perform an installation or removal procedure on standard desktops, laptops, and Tablet PCs one step at a time, to ensure that the technician clearly understands the procedure and is ready to perform the procedure using an actual assembly. Three-dimensional models are used in the simulation to enable the technician to feel as though they were performing the removal or installation procedure using the actual assembly. Using three-dimensional simulations technicians can perform, view, and understand the procedure using a three-dimensional view. In installation, parts are dragged from a bin to create an assembly. In removal, parts are dragged from the assembly into a bin. In either case if a wrong part is installed or removed, an alert box is displayed on the screen preventing the technician from proceeding until the correct part is installed or removed.

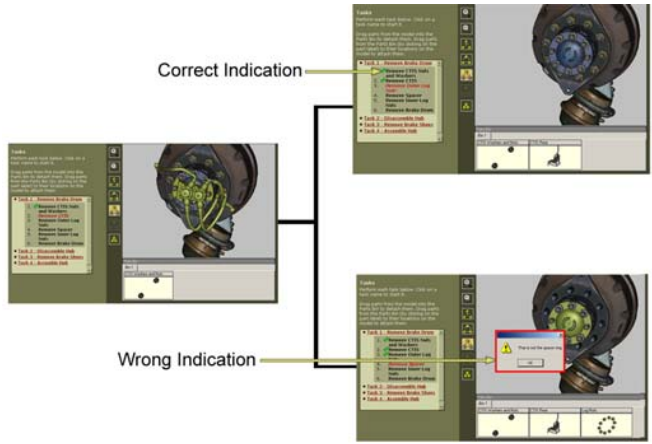

Figure 5.   Practice module

## VII.   SUCCESSFUL DEPLOYMENT

### A.   Turbojet Engines

Ineffective training to technicians that are geographically distributed has resulted in improper troubleshooting procedures being carried out on Turbojet Engines resulting in reliability of the engine being compromised. Technicians are now being trained using interactive 3D simulations of the engine explaining its description, operation, maintenance and troubleshooting procedures resulting in an estimated saving of $1.5 Million with an improved maintenance turn-around-time. Technicians now are able to practice these procedures on standard desktops, laptops, and Tablet PCs eliminating geographic barriers and imparting a high standard of training.

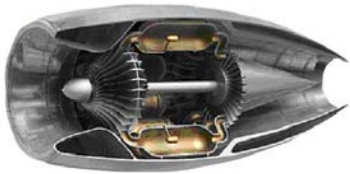

Figure 6.   Turbojet Engine

### B.   Anti-tank and Anti-personel mines

There is a difficulty in training soldiers in the Army on handling and disposal of both anti-tank mines and anti-personnel mines.  Anti-tank mines are large and heavy (usually weighing more than 5 kilos), triggered by heavy vehicles such as tanks. These mines contain enough explosives to destroy the vehicle that runs over them.  Anti-personnel mines are smaller and lighter (weighing as little as 50 grams), triggered much more easily and is designed to wound people. It is critical that these soldiers have access to technical information about the landmine, details regarding safe handling and its disposal. Creation of 3D simulations of landmines that allow soldiers to view its details of its parts, watch safety procedural animations and interact with them resulted in soldiers having greater understanding and knowledge of landmines they encounter.

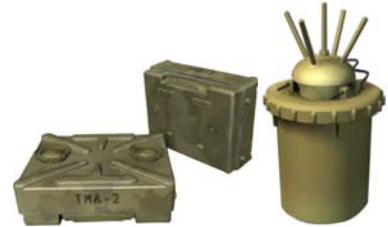

Figure 7.   Anti-tank and Anti-personel mines

### C.   M79 Grenade Launcher

It had been identified by the Army that a lack of access to M79 Grenade Launchers during familiarization trainings had resulted in deployment of soldiers with limited knowledge and experience levels. To overcome this hurdle 3D-enabled M79 Grenade Launchers Virtual Task Trainers were provided simulating the single-shot, shoulder-fired, break open grenade launcher, which fires a 40x46mm grenade. Now soldiers are able to receive familiarization training regardless of geographic barriers or lack of access to weapons.

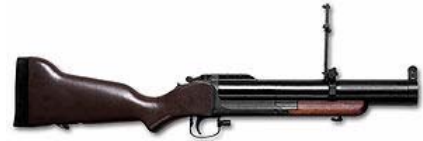

Figure 8.   M79 Grenade Launcher

### D.   Black Shark Torpedo

Black Shark torpedo is designed for launching from submarines or surface vessels. It is meant to counter the threat posed by any type of surface or underwater target. Due to the fast pace of operations, Navy technicians received little to no training on Black Shark torpedo. This has resulted in improper operating procedures and preventative maintenance checks. Web-enabled 3D simulations have been developed allowing technicians to have hands-on practice anytime and anywhere along with familiarization to parts, maintenance procedures and repair tasks. This has resulted in technicians showing a level of interest using 3D simulations compared to existing course materials like PowerPoint presentations, manuals and videos.

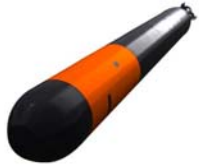

Figure 9.   Black Shark Torpedo

### E.   Phoenix Missile

Technicians were constantly facing operational difficulties concerning Phoenix Missile due to the inability to demonstrate the operation of its internal components. Phoenix Missile is a long-range air-to-air missile. Interactive 3D simulation demonstrating its internal components along with functioning was developed allowing technicians to view parts information, rotate and cross-section of the Phoenix Missile.

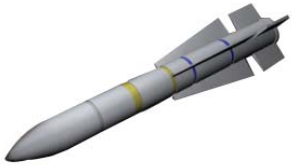

Figure 10.  Phoenix Missile

## VIII.   CASE STUDIES

A case study to find out the effort saved using interactive 3D was carried out on the following installations:

- Hydraulic Pump
- Hydraulic Reservoir
- High Pressure Filter
- Anti-Skid Control Valve
- Quick Disconnect Coupling Suction

TABLE I.          USING TRADITIONAL TOOLS LIKE VIDEO FOR TRAINING

| Installation | Time spent in training a trainee using video clips | Actual time spent to complete the installation by a trainee |
|---|---|---|
| Hydraulic Pump | 3 hours | 30 minutes |
| Hydraulic Reservoir | 4 hours 30 minutes | 1 hour |
| High Pressure Filter | 3 hours | 45 minutes |
| Anti-Skid Control Valve | 2 hours 30 minutes | 45 minutes |
| Quick Disconnect Coupling Suction | 2 hours 15 minutes | 30 minutes |

TABLE II.          USING INTERACTIVE 3D FOR TRAINING

| Installation | Time spent in training a trainee using Interactive 3D | Actual time spent to complete the installation by a trainee |
|---|---|---|
| Hydraulic Pump | 2 hours | 20 minutes |
| Hydraulic Reservoir | 3 hours | 45 minutes |
| High Pressure Filter | 1 hours 30 minutes | 30 minutes |
| Anti-Skid Control Valve | 1 hours 30 minutes | 30 minutes |
| Quick Disconnect Coupling Suction | 1 hours 30 minutes | 20 minutes |

TABLE III.          PERCENTAGE OF EFFORT SAVED IN USING INTERACTIVE 3D FOR TRAINING

| Installation | Time spent in training a trainee using Interactive 3D | Actual time spent to complete the installation by a trainee |
|---|---|---|
| Hydraulic Pump | 33.3% | 34% |
| Hydraulic Reservoir | 33.3% | 25% |
| High Pressure Filter | 50% | 33.3% |
| Anti-Skid Control Valve | 40% | 33.3% |
| Quick Disconnect Coupling Suction | 50% | 34% |

## IX.   CONCLUSION

The metrics collected and analyzed during the implementation of interactive 3D demonstrates the following benefits in maintenance

- Reducing the time spent in training by 30% or more
- Reducing the time spent in performing the installation by a trainee by 25% or more

In conclusion interactive 3D reduces the amount of time spent in hands-on training on real equipment, protects trainees from injury and equipment from damage when performing procedures that are hazardous. It provides an opportunity for technicians to study the internal components of equipments and this training can be provided to technicians that are geographically distributed.